\documentclass[12pt]{article}
\usepackage[latin1]{inputenc}
\usepackage{amssymb}

\newcommand{\tr}{\hbox{tr}}
\newcommand{\Str}{\hbox{Str}}

\begin{document}

\title{Supersymmetric and R-symmetric vacua in Wess-Zumino models}

\author{
\small{\hfill LPTENS-07/38}\\
S\'ebastien Ray\\
\it\normalsize	Laboratoire de physique th\'eorique de l'\'Ecole normale
	sup\'erieure\\
\it\normalsize	24 rue Lhomond, 75231 Paris Cedex 05, France}

\maketitle

\abstract{In the context of supersymmetric Wess-Zumino models with an R 
symmetry, we find some simple conditions on the R-charge content of the 
theory that imply the presence or absence of supersymmetric and 
R-symmetric vacua. The main result of this work is that the comparison 
between the number of R-charge 0 and R-charge 2 superfields is essential 
to the properties of the model as regards symmetry breaking. We also 
study possible exceptions to the Nelson-Seiberg theorem ---finding that 
there \emph{are} supersymmetric vacua that break R symmetry in generic 
models--- and the spontaneous breaking of R symmetry in 
supersummetry-breaking vacua, with some insight on the Coleman-Weinberg 
one-loop potential.}

\small

\section{Introduction}

Supersymmetry breaking is coming back in the foreground of theoretical 
physics, which is hardly surprising when it is about to become 
experimentally testable. Among recent theoretical achievements, 
attention was brought \cite{Intriligator:2006dd} on the possibility of 
circumventing part of the strong constraints put on 
supersymmetry-breaking models \cite{Witten:1982df} by considering that 
supersymmetry could be broken in a metastable vacuum, thereby 
authorizing a supersymmetric state elsewhere in the phase space.

In that context, the models of Wess and Zumino \cite{Wess:1973kz}, 
formed with only chiral superfields, have gained renewed interest, since 
they offer all the interesting features of supersymmetry breaking 
(dynamic breaking, metastable vacua...) and their simplicity makes them 
easy to handle. They are thus good toy models in our search for a 
realistic supersymmetry breaking mechanism. These models are moreover of 
some interest in their own right as they could be used as hidden sector 
initiating a supersymmetry breaking that could then be propagated to the 
standard model by some mediation. Another source of interest is the fact 
that some more elaborate ---and realistic--- models, such as the SQCD 
model of \cite{Intriligator:2006dd}, have a Wess-Zumino low-energy 
limit.

\bigskip

\emph{Motivations}

\bigskip

Since supersymmetry is not observed in low-energy physics, a 
supersymmetric theory of particle physics should provide for a 
supersymmetry-breaking process. This is not so easy to achieve: the 
constraints on supersymmetry were summarized by Witten in 
\cite{Witten:1982df} and a landmark article of Nelson and Seiberg 
\cite{Nelson:1993nf} showed that the presence of an R symmetry, which is 
a symmetry under which the supersymmetry generator carries a charge, was 
an essential ingredient for breaking supersymmetry in a generic model.

But experimental constraints require that R symmetry should also be 
broken in nature. The sole allowed remnant of it would be R parity (see 
\cite{Barbier:2004ez} for a review). It is therefore necessary to have 
an understanding of the relationship between R symmetry-breaking and 
supersymmetry-breaking. The R-symmetric Wess-Zumino models provide a 
simple and relevant frame for studying both mechanisms.

\bigskip

The purpose of this note is to emphasize some necessary or sufficient 
conditions on the R-charges of the superfields with respect to the 
existence of supersymmetric or R-symmetric vacua. The first part, which 
owns much to \cite{Nelson:1993nf}, deals with supersymmetric vacua, 
whereas the second one, studying supersymmetry breaking, will rely in 
particular on the excellent work recently done in \cite{Shih:2007av}, 
while bringing precisions to some of its demonstrations and completing 
it in some unexplored directions.

The (partial) conclusions of this paper as to the existence of different 
types of vacua can be summarized in the following table:

\bigskip

\begin{table}
\begin{center}
\begin{tabular}{|c|c|c|}
\cline{2-3}
\multicolumn{1}{c|}{} & SuSy & SuSy-B \\ \hline
RSy & $n_0\geq n_2$: $\exists$ vacuum & $n_0\geq n_2$: NO \\
& $n_0<n_2$: NO & $n_0<n_2$: possible\\ \hline
RSy-B & no (with exceptions) & possible\\ \hline
\end{tabular}
\end{center}
\caption{\label{conclusions} Overview of symmetries and symmetry 
breakings in R-symmetric Wess-Zumino models.}
\end{table}

\bigskip

\emph{Conventions}

\bigskip

Throughout this paper, we shall consider N=1 Wess-Zumino models\footnote{%
 The models with chiral superfields are sometimes called \'O Raifeartaigh 
 models, in reference to the supersymmetry-breaking model found by 
 Lochlainn \'O Raifeartaigh in \cite{Raifeartach:1975pr}. For historical 
 exactness, I shall reserve that name for the specific 
 supersymmetry-breaking model of 1975 and name the general model of 
 chiral superfields introduced in 1973 \cite{Wess:1973kz} after its 
 authors Wess and Zumino.
 }
involving $N$ chiral superfields\footnote{%
 The notation $\phi^a$ shall denote either the chiral superfield or its
 scalar component, depending on the context.
 }
$\phi^a$ with canonical K\"ahler potential $K=\phi_a^\dagger \phi^a$ and 
a superpotential $W(\phi)$, which is a holomorphic function of the 
superfields.

This model shall admit an R symmetry $U(1)_R$, meaning that each field has 
an R-charge so that $W(\phi)$ has R-charge $2$. The supersymmetric action 
writes, in terms of the superfields:

$$
S=\int d^4x \left[ \int d^2\theta d^2\bar\theta K + \int d^2\theta W + 
h.c. \right]
$$

We shall call \emph{vacuum} a locally stable state, \emph{i.e.} a state 
where the energy is locally minimal. A vacuum can be \emph{metastable} 
if a state with lower energy is available elsewhere in the phase space, 
in which case a disintegration is possible through non-perturbative 
effects. If the vacuum is a global minimum of the energy, it is a stable 
(or absolutely stable) vacuum.

We shall call \emph{degenerate} a vacuum that belongs to a set of vacua 
with the same energy forming a (continuous) submanifold of the field 
space. The dimension of that submanifold will be the \emph{degeneracy} 
of the vacuum. If a vacuum breaks an ordinary symmetry or an R symmetry, 
it is necessarily degenerate, as stated by the Goldstone theorem.

We shall use the word \emph{symmetry} to design both R symmetry and 
supersymmetry. A \emph{symmetric} vacuum is a vacuum that is invariant 
under the symmetry in question, as opposed to a \emph{symmetry-breaking} 
vacuum. Symmetry breaking is not equivalent to the absence of symmetric 
vacuum: there can be several vacua, some being symmetric and some 
symmetry-breaking.

Although a well-known result of supersymmetry is that the superpotential 
is not renormalized, we shall call \emph{renormalizable} a superpotential 
that leads to a renormalizable potential for the bosons, meaning that it 
is polynomial of degree at most three in the superfields.

Finally, a \emph{generic} model is a model in which no free parameter, 
or combination of free parameters, has a special value (\emph{e.g.} 
zero) unless it is required by some symmetry, for instance R symmetry. 
In the set of all models with a given symmetry and a given field 
content, \emph{nearly all} models are generic and therefore share the 
properties ascribed to generic models in this paper.

\section{Supersymmetric vacua}

In this section, we shall consider generic models, with the only 
restriction that the superpotential be an integer series in the fields 
around an R-symmetric point in field space.

\subsection{Supersymmetric R-symmetric vacua}

Let us consider an R-symmetric state. This means that all quantities 
carrying a non-zero R-charge must have zero expectation value. The 
derivative $\partial_a W$ of the superpotential with respect to a field 
$\phi^a$ of R-charge $R_a$ has an R-charge $2-R_a$, so that its 
expectation value must be zero if $R_a\neq2$. If no field has R-charge 
$2$, then $\partial_a W$ must be zero in any R-symmetric configuration. 
From this we draw a first conclusion, valid for non-generic as well as 
generic models:

\bigskip

\emph{I. In any model with no field of R-charge $2$, there is a 
supersymmetric vacuum; in fact, all R-symmetric states are supersymmetric 
vacua.}

They are therefore degenerate if there are R-neutral fields, which are 
then massless.

\bigskip

A simple example is the generic renormalizable model with two fields of 
R-charges 0 and 1:

\begin{equation}
W=\frac{1}{2}\phi_1^2 f(\phi_0).
\end{equation}

The supersymmetry conditions yield only $\phi_1=0$, which is exactly the 
R symmetry condition; the potential is degenerate along the R-neutral 
field $\phi_0$.

\bigskip

As a corollary, if we suppose that there exists a non-supersymmetric 
vacuum elsewhere in field space, it cannot be R-symmetric:

\bigskip

\emph{I'. If there is no field of R-charge $2$, supersymmetry breaking 
implies R symmetry breaking.} A symmetry-breaking vacuum is always 
metastable, since the R-symmetric (supersymmetric) vacua have less 
energy.

\bigskip

We can generalize this conclusion by using genericity: consider a model 
where there are fields of R-charge $2$ ---let us call $n_2$ the number 
of these fields--- and, similarly, $n_0$ fields of R-charge zero. The 
equations that must be solved for an \emph{R-symmetric} state to be 
supersymmetric can be written:

$$ \partial_{(2)} W(\phi_{(0)})=0,$$

where $(2)$ represents the fields of R-charge $2$ (all $\partial_iW$ are 
automatically zero if $R_i\neq2$) and $\phi_{(0)}$ the R-neutral 
fields, which are the only fields that are allowed to have non-zero 
value. We thus have a set of $n_2$ equations with $n_0$ variables: this 
admits a solution for a generic choice of parameters if and only if $n_0 
\geq n_2$. More precisely, there will generically be a 
$(n_0-n_2)$-dimensional set of solutions. A second conclusion can 
therefore be reached:

\bigskip

\emph{II. When a model contains $n_2$ fields of R-charge $2$ and $n_0$ 
R-neutral fields, there generically exists a R-symmetric supersymmetric 
vacuum if and only if $n_0\geq n_2$.}

If $n_0>n_2$, that solution is degenerate in $(n_0-n_2)$ directions.

\bigskip

Note that if an additional (ordinary) symmetry of the fields is present, 
then the function $W$ is not ``generic'' anymore in our sense and the 
conclusion could be modified, with the further complication that R 
symmetry is not uniquely defined in that context, since any ordinary 
charge can be added to $R$ to form a new R symmetry. It should be 
possible, with some caution, to find a generalization of conclusion II. to 
cases with additional symmetries, but we shall limit ourselves in this 
paper to the case where the only symmetries are $N=1$ supersymmetry 
and $U(1)_R$ symmetry.

\subsection{Supersymmetric R symmetry-breaking vacua}

We can add to these results the well-known conclusion of Nelson and 
Seiberg, who showed in \cite{Nelson:1993nf} that the existence of a 
supersymmetric vacuum in a generic model could only be avoided in presence 
of an R symmetry that should be spontaneously broken. In other words:

\bigskip

\emph{IIIa. There generically exists no R symmetry-breaking 
supersymmetric vacuum.}

\bigskip

A supersymmetric state must therefore be R-symmetric in a generic model. 
Let us rapidly recall the demonstration leading to that conclusion: 
outside R-symmetric states, we can always choose a field $X$ of R charge 
$R\neq0$ with a non-zero expectation value. All $(N-1)$ other fields 
$\phi^i$ can then be written in terms of R-neutral fields $\varphi^i$ as 
$\phi^i \equiv \varphi^i X^{R_i/R}$, so that:

\begin{equation}
\label{rsupot}
W(\phi^a)=X^{2/R}f(\varphi^i)
\end{equation}

for some function $f$. The supersymmetry conditions can be written in 
terms of the $\varphi^i$s only, as:

\begin{equation}
\label{susy-rb}
\partial_i f=0,\; f=0,
\end{equation}

which makes $N$ equations for $N-1$ variables, a generically unsoluble 
system.

\bigskip

\emph{Exceptions to Nelson-Seiberg}

\bigskip

An important \emph{caveat} must be added here: as already noted in 
\cite{Intriligator:2007cp}, there are exceptions to that rule. They arise 
from the fact that the function $f$ is not fully generic. Indeed, a 
general superpotential must be a (locally) analytic function of the 
superfields, so that Taylor developping $f$ in expression (\ref{rsupot}) 
should lead to an expression where the powers of $X$ are non-negative 
integers.

It is not easy to formulate a general condition on the R-charges of the 
theory for the existence of R symmetry-breaking supersymmetric vacua. 
Still, we can find a whole class of exceptions if we search for models 
in which an R symmetry-breaking vacuum is constructed from an 
R-symmetric supersymmetric vacuum (which we know exists if $n_2\leq 
n_0$) that is degenerate in an R symmetry-breaking direction. This is 
quite a natural condition since, if the equations (\ref{susy-rb}) have 
a solution $\{\varphi^i\}$, it will lead, not to an isolated 
supersymmetric vacuum, but to a whole line of degenerate supersymmetric 
vacua $(X, \varphi^i X^{R_i/R})$, where $X$ can take any (complex) 
value. We only impose that this line contains an R-symmetric state, 
which shall necessarily be at $X=0$: that is to say that all fields 
$\phi^i$ with $R_i/R<0$ should be zero.

The condition that such a line of supersymmetric vacua exists is 
therefore:

$$
\exists \varphi^b_+, \phi_{(0)}; \forall X, \forall a, 
\partial_a W (X, \varphi^b_+ X^{R_b/R}, \phi_{(0)}, \phi_-=0)=0, 
$$

where $\phi_{\pm}$ represents fields with R charge of the same (opposite) 
sign as $R$. Expanding that equation in $X$ around the R-symmetric state 
$X=0$ leads to:

$$
\exists \varphi_+^b, \phi_{(0)}; \forall \alpha>0, \forall a, 
\sum_{i,j ; \{b_k\}} \frac{1}{i!j!} \left(\prod_{k=1}^i 
\varphi_+^{b_k}\right) \partial^i_{b_1 \ldots b_i} \partial_X^j \partial_a 
W (\phi_{(0)})=0
$$

where the set of fields $\{b_k\}$ should satisfy the condition $\sum 
R_{b_k} = (\alpha-j) R$. As that equation must hold in an R-symmetric 
state it is trivially satisfied if the left-hand term has a non-zero R 
charge: as that R charge is $(2-\alpha R-R_a)$ only the derivatives with 
respect to fields of R charge $(2-\alpha R)$ yield equations for each 
$\alpha$.

For each $\alpha>0$, we thus have $n_{2-\alpha R}$ equations, one for 
each field of the theory carrying R charge $(2-\alpha R)$ ; these 
equations bear on the $n_0$ R-neutral fields $\phi_{(0)}$ \emph{and} on 
the $\varphi_+^b$ corresponding to the R-charged fields $\phi^i_+$ with 
R-charges of the same sign as $X$ and which can somehow be added up to 
$\alpha R$, $(\alpha-1)R$,... This system is generically solvable if and 
only if all subsets of equations bear on a superior number of variables. 
This can be written:

\begin{equation}
\label{NSconditions}
\left\{
\begin{array}{rcl}
n_2 \leq n_0 && \\
\forall E\subset \mathbb{R}^+, E\notin\{\emptyset,\{0\}\}, \sum_{\alpha 
\in E} n_{2-\alpha R} &\leq& n_0 + \sum_{r\in F_E} \bar n_{r},
\end{array}\right.
\end{equation}

where $F_E \equiv \{r; r/R>0, r/R$ summable up to $\alpha, \alpha-1,...$ 
for some $\alpha\in E\}$. $\bar n_r$ is equal to $n_r$ except for $\bar 
n_R=n_R-1$, the field $X$ being absent as a variable in the system.

Note that if all R-charges in the model are positive (which implies that R 
charges $R>2$ cannot contribute to the superpotential) the conditions are 
never met, since there are as many equations as superfields whereas there 
is one variable ($X$) less (in the above expression, take 
$E=\mathbb{R}^+$).

As this condition is not really intuitive, some examples given in Annex 
I. can help to see the point.

\bigskip

We can now write the partial conclusion:

\bigskip

\emph{IIIb. There are exceptions to 3a. One class of exceptions is given 
by the models satisfying conditions} (\ref{NSconditions}). \emph{These 
models have R symmetry-breaking supersymmetric vacua, which are 
degenerate. }

\bigskip

Note that there are models in which other kinds of lines of R 
symmetry-breaking supersymmetric vacua are present, which do not contain 
an R symmetric state. An example is given by a three-field theory 
$(\phi_2,X_3,Y_{-3})$: a generic superpotential will 
write

\begin{equation}
\label{exception}
W=\phi f(\phi^3 Y^2,XY),
\end{equation}

and there is a line of supersymmetric vacua for $\phi=0$, $f(XY,0)=0$, 
breaking R symmetry, whereas $n_2>n_0$ insures that the R-symmetric state 
$\phi=X=Y=0$ is not supersymmetric.

\bigskip

It is interesting to notice here that \cite{Nelson:1993nf} quotes an 
older article on supersymmetry breaking \cite{Affleck:1983vc}, where it 
is stated that a condition for dynamical supersymmetry breaking is that 
the scalar potential have no flat direction at infinity. This condition 
is clearly not met in the cases we have just found since the freedom in 
$X$ implies that these models have a whole line of supersymmetric vacua, 
extending to $X\rightarrow\infty$. Thus, although we did find exceptions 
to the general conclusions of \cite{Nelson:1993nf}, they do not 
contradict the more fundamental principles of \cite{Affleck:1983vc}.

\subsection{First conclusions}

From these first results we can classify the Wess-Zumino R-symmetric 
models in two groups:

\begin{itemize}

\item \emph{When $n_2>n_0$, there generically exists no supersymmetric 
vacuum at all.}

The original \'O Raifeartaigh model \cite{Raifeartach:1975pr} falls in 
this category (it has three fields of R-charges $2$, $2$ and $0$), as well 
as the Shih model (R-charges $-1$, $1$, $2$ and $3$) introduced in 
\cite{Shih:2007av}.

There are rare exceptions to this rule (\ref{exception}). In those cases 
the supersymmetric vacua are degenerate and break R symmetry.

Whether there exists supersymmetry-breaking vacua, or indeed any vacuum 
at all, depends on the model. A very simple model will be of some use to 
illustrate this point: a generic renormalizable superpotential using 
only two fields $X$ and $\phi$ of R-charges $2$ and $-2$ is:

\begin{equation}
\label{model2-2}
W=\xi X + \frac{1}{2}\lambda X^2 \phi,
\end{equation}

where $\lambda$ and $\xi$ can be chosen real positive by field 
redefinition. This model, since $n_2>n_0$, should break both R symmetry 
and supersymmetry. Its tree-level potential is:

$$ V_0=\xi^2 +2\lambda\xi \Re(X\phi) + \lambda^2 |X|^2|\phi|^2 + 
\frac{1}{4}\lambda^2 |X|^4.$$

The only extremum is at the R-symmetric state $X=\phi=0$, but this 
extremum is not a minimum, the direction $\delta X=-\delta \phi$ being 
obviously tachyonic. In fact, this model is an incongruous case where 
there is no vacuum at all, only a runaway $\phi\rightarrow\infty$, 
$X=-\xi/(\lambda \phi)$. It was signalled in \cite{Witten:1981kv} and 
more recently in \cite{Ferretti:2007ec}; the properties remain the same 
if we remove the renormalizability condition\footnote{That model could 
have interesting applications in the mass hierarchy problem, since the 
runaway is in fact stabilized by the inclusion of (super)gravity, 
thereby generating two mass scales, one naturally small and one 
naturally large. This would deserve a more thorough investigation than 
can be included in the frame of the present work.}.

\item \emph{When $n_2\leq n_0$ there generically exists supersymmetric 
vacua, which are R-symmetric, with a degeneracy of order $(n_0-n_2)$.}

There can be R symmetry-breaking supersymmetric vacua as well, for 
instance if condition (\ref{NSconditions}) is met.

The simplest non-trivial example is the model with two fields of 
R-charges $0$ and $2$. The generic superpotential is:

\begin{equation}
\label{model20}
W=\phi_2f(\phi_0)
\end{equation}

There are supersymmetric vacua, located at $\phi_2=0$ and $f(\phi_0)=0$, 
which are also R-symmetric and non-degenerate, as expected since 
$n_0-n_2=0$.

\end{itemize}

To look at things from the other side, global supersymmetry breaking (in 
the sense of the absence of any supersymmetric vacuum) in a generic WZ 
model requires not only the presence of an R symmetry 
\cite{Nelson:1993nf}, but that there be (strictly) more fields with 
R-charge $2$ than R-neutral fields ---though this is, again, only a 
necessary condition.

\section{Supersymmetry-breaking R-symmetric vacua}

A physical model must not only account for supersymmetry breaking, but 
also for R symmetry breaking. Therefore, having studied supersymmetric 
vacua (or the absence thereof), we can now look for conditions under 
which R-symmetric supersymmetry-breaking vacua exist. Most of the 
results in this section shall be limited to renormalizable 
superpotentials.

\subsection{General results}

Supersymmetry-breaking vacua are less simple to study since they cannot 
be characterized by the beautiful, simple, necessary and sufficient 
condition $\partial W=0$. Still, some things can be said about them when 
they are imposed to be R-symmetric.

As we showed in \cite{Ray:2006wk}, a supersymmetry-breaking vacuum 
implies a tree-level degeneracy in the direction of the expectation 
value of $\partial W^\dagger$, which must of course be non-zero. This is 
possible in an R symmetric vacuum only if there is a field with R-charge 
$2$ (which we already knew from conclusion I.).

If we call $X$ the R-charge $2$ field given by the direction of the 
expectation value of $\partial W^\dagger$, we showed in 
\cite{Ray:2006wk} that for any strictly positive integer $k$, the 
expectation value of $\partial_a \partial^k_X W$ is zero in the vacuum.

\bigskip

For $k=1$, this is simply the equation of extremization of the 
potential. As it is automatically satified for fields $\phi^a$ with a 
non-zero R-charge, the vacuum is only characterized by $n_0$ equations 
bearing on the $n_0$ remaining free variables $\phi_{(0)}$ of the space 
of R-symmetric states. This equation generically has at least a 
solution and fixes the values of the R-neutral fields.

For $k>1$, as we have no more freedom on the $\phi_{(0)}$'s left, the 
additional condition $\partial_a \partial^k_X W=0$ can only be true in a 
generic model if its R-charge is always non-zero, that is, if there exists 
no field of R-charge $2-2k$:

\bigskip

\emph{IVa. A generic model with a field of R-charge $2$ admits a  
supersymmetry-breaking R-symmetric vacuum only if it contains no field 
of R-charge $-2, -4, -6\ldots$}

\bigskip

But if the model is imposed to be renormalizable, then the superpotential 
is at most trilinear and we need only consider the case $k=2$, from which 
we conclude:

\bigskip

\emph{IVb. A necessary condition for a generic renormalizable model to 
admit a supersymmetry-breaking R-symmetric vacuum is to contain at least a 
field of R-charge 2 and no field of R-charge $-2$.}

\bigskip

In other words, a generic model with a field of R-charge $-2$ will, if 
the Nelson-Seiberg result III. holds, either preserve both R symmetry 
and supersymmetry ($n_0\geq n_2$) or break both ($n_0<n_2$). An instance 
of the former is the abovementioned $(2,0)$ model (\ref{model20}), 
whereas an instance of the latter is the $(-2,2)$ model 
(\ref{model2-2}).

\bigskip

Once the necessary condition IVb. is verified, the (renormalizable) 
superpotential can be put in the form:

\begin{equation}
\label{superpot}
W=\xi X + \frac{1}{2}(\mu_{ij}+X \lambda_{ij})\phi^i \phi^j + 
\frac{1}{6}\gamma_{ijk}\phi^i\phi^j\phi^k.
\end{equation}

The extremum of the tree-level potential is then at $\phi_i=0$ for any 
$X$. Now in order for the R-symmetric state $X=0$ to be a vacuum, two more 
conditions should be simultaneously met:

\begin{itemize}
\item the directions $\phi_i$ should be non-tachyonic,
\item as well as the one-loop-generated potential on $X$.
\end{itemize}

This is a non-trivial problem, which we study in the next two sections.

\subsection{Local stability of the R-symmetric states}

A renormalizable R-symmetric Wess-Zumino model with a field $R=2$ and no 
field $R=-2$ generically has an R-symmetric extremum around which the 
superpotential can be written in the form (\ref{superpot}). R symmetry 
imposes the relations $R(X)=2$, $\mu_{ij}\neq0 \Rightarrow 
R(\phi_i)+R(\phi_j)=2$, $\lambda_{ij}\neq0 \Rightarrow 
R(\phi_i)+R(\phi_j)=0$ and similar relations for $\gamma_{ijk}$. The 
mass matrix of the bosons $\phi_i$ is given by:

$$
M_0^2=\left(\begin{array}{cc}
\mu^\dagger\mu & \xi\lambda^\dagger \\
\xi\lambda & \mu \mu^\dagger\end{array}\right).
$$

(We can always choose $\xi$ real positive.) A necessary condition for 
the extremal state to be a local minimum is that this matrix be 
positive, which we can write:

\begin{equation}
\label{positive}
\forall \psi_1, \psi_2, \|\mu\psi_1\|^2 + \|\mu^\dagger \psi_2\|^2 
+2\xi\Re(\psi_2^\dagger \lambda\psi_1) \geq 0.
\end{equation}

We can see that only the matrices $\mu$ and $\lambda$ come in that 
relation. As an R-charge $R$ can only be connected by $\mu$ to an 
R-charge $2-R$ and by $\lambda$ to $-R$, the R-charges of the $\phi$ 
fields can be organized in sequences of R-charges alternatively related 
by $\mu$s and $\lambda$s:

$$
\begin{array}{ccccccccccccc}
&&R-2&&&&R&&&&R+2&& \\
&\mu\diagup&&\lambda\diagdown&&\mu\diagup&&\lambda\diagdown&&\mu\diagup&
&\lambda\diagdown&\\
\ldots&&&&2-R&&&&-R&&&&\ldots
\end{array}
$$

There can be independent series, which will perhaps be related by 
$\gamma_{ijk}$, but that is of no importance for the mass calculation. 
If there is an ordinary symmetry in addition to R symmetry, the 
sequences should be separated according to the charges under that 
symmetry.

In a model with a finite number of fields, each of these sequences must 
end somewhere on both sides. But if it ends after a $\lambda$ link we 
shall have fields of R-charge $R$ linked by $\lambda$ to fields $-R$ but 
with no R-charge $2-R$ to be linked with by $\mu$. In that situation, 
taking in relation (\ref{positive}) a $\psi_1$ of R-charge $R$, we have 
$\mu\psi_1=0$, so that the relation can only be true if 
$\lambda_{-R,R}=0$, where $\lambda_{-R,R}$ represents the submatrix of 
$\lambda$ linking $R$ and $-R$. But in absence of a symmetry justifying 
this nullity, it requires fine tuning, so that in a generic model this 
hypothesis must be excluded: if the sequence of R-charges ends with a 
$\lambda$ relation, then there is a tachyonic direction. A necessary 
condition for having a vacuum is therefore that the sequence of 
R-charges end with a $\mu$ relation on both sides.

\bigskip

We can still draw a further condition from genericity: if, in a 
sequence, two sets of $\phi$ fields, one with R-charge $R$, the other 
with R-charge $2-R$, linked by $\mu$, have a different number of fields 
---say for instance that $n_{2-R}<n_R$---, then there must exist some 
$\psi_1$ of R-charge $R$ for which $\mu\psi_1=0$. In that case, again, 
(\ref{positive}) implies that $\lambda\psi_1$ is zero too, which again 
requires fine tuning unless there exists no field of R-charge $-R$. Thus 
two R-charges linked by $\mu$ inside a sequence must have the same 
number of fields. Only the R-charges at both extremities can be an 
exeption, with a possible greater number of fields.

(Note that for $R=2$, the $X$ field is \emph{not} counted in $n_2$.)

We can summarize this in the following drawing:

$$
\begin{array}{ccccccccccccc}
&&2-R'&&&&\ldots&&R-2&&&&R \\
&\mu\nearrow&&\lambda\diagdown&&\diagup\diagup\mu&\ldots
&\diagup\diagup\mu&&\lambda\diagdown&&\mu\swarrow&\\
R'&&&&R'-2&&\ldots&&&&2-R&&
\end{array}
$$

where $\nearrow$, $\swarrow$ and $\diagup\diagup$ indicate a relation 
between two sets with a superior (or equal), inferior (or equal) or 
equal number of fields.

\bigskip

There are two particular cases: R-charges $0$ can be 
self-$\lambda$-coupled and R-charges $1$ self-$\mu$-coupled ---if there 
is an additional ordinary symmetry, this is only possible if the fields 
are neutral---, so that we can have semi-sequences starting at $R=0$ or 
$R=1$. The one starting at $R=0$ must almost immediately stop because 
$R=-2$ fields are prohibited. The following sequences are therefore 
allowed:

$$
\begin{array}{ccc}
&&2\\
&\mu\swarrow&\\
\lambda\subset 0&&
\end{array};\quad\quad\quad\quad\quad
\begin{array}{cccccc}
\mu\subset 1&&&&3&\ldots\\
&\lambda\diagdown&&\diagup\mu&&\ldots\\
&&-1&&&\ldots
\end{array}
$$

The original \'O Raifeartaigh model is of the first form and the Shih 
model of the second.

\bigskip

We can therefore write the following necessary condition for an 
R-symmetric model to be able to have an R-symmetric supersymmetry-breaking 
vacuum without fine tuning:

\bigskip

\it
V. A generic renormalizable model can have an R-symmetric 
supersymmetry-breaking vacuum only if: 

\begin{itemize}

\item it has a field $X$ with $R=2$ and no field with $R=-2$;

\item all other fields have R-charges which can be arranged in one or 
several of the following sequences or their subsequences:
\end{itemize}
\rm

$$
\begin{array}{ccc}
&&2\\
&\mu\swarrow&\\
\lambda\subset 0&&
\end{array};\quad\quad\quad\quad\quad
\begin{array}{cccccc}
\mu\subset 1&&&&3&\ldots\\
&\lambda\diagdown&&\diagup\mu&&\ldots\\
&&-1&&&\ldots
\end{array};
$$

$$
\begin{array}{ccccccccccc}
&&2-R'&&\ldots&&R-2&&&&R \\
&\mu\nearrow&&\lambda\diagdown&\ldots
&\diagup\diagup\mu&&\lambda\diagdown&&\mu\swarrow&\\
R'&&&&\ldots&&&&2-R&&
\end{array}.
$$

\bigskip

We must note that these, again, are only necessary conditions. There 
cannot be a complete characterization of R-symmetric vacua, since it is 
obvious from the form of (\ref{positive}) that the $\lambda$ coefficients 
can always be chosen large enough to destabilize an extremum in a $\phi$ 
direction and that, given condition V. and for generic $\mu$, they can 
always be chosen small enough to insure stability.

\bigskip

An important consequence of conclusion V. is that R-symmetric 
supersymmetry-breaking vacua generically cannot exist if $n_0>(n_2-1)$ 
(here $n_2$ includes the $X$ field), meaning $n_0\geq n_2$, which is 
exactly the condition of existence of supersymmetric vacua. This means 
that a generic model cannot have both a supersymmetric and a R-symmetric 
supersymmetry-breaking vacuum ---with possible exceptions in 
Nelson-Seiberg-violating models.

\bigskip

So much for the $\phi$ directions; we must now study the tree-level 
degererate $X$ direction, which is the object of the last section.

\subsection{Pseudomodulus stability}

As shown in Annex II, the pseudomodulus $X$ is given a mass by one-loop 
effects:

\begin{eqnarray*}
m_X^2 &=& m_1^2-m_2^2\\
m_1^2 \geq 0,\; =0 &\Leftrightarrow& \lambda=0\\
m_2^2 \geq 0,\; =0 &\Leftrightarrow& \forall k\geq0, \lambda 
(\mu^\dagger \mu)^k \mu^\dagger \lambda = 0.
\end{eqnarray*}

There are thus two concurring effects and the resulting squared mass 
can be either positive or negative. We can force it to be positive by 
imposing $m_2=0$.

From the graphical point of view seen above, that matrix 
$\lambda(\mu^\dagger \mu)^k \mu^\dagger \lambda$ corresponds to 
following one $\lambda$ line, then doing an odd number of comings and 
goings on a $\mu$ line ---which is equivalent to following just one 
$\mu$ line---, then following again a $\lambda$ line. This cannot be 
always zero without fine tuning unless such a travel is indeed 
impossible given the available R-charges. In other words, the sequences 
described in condition V. must be such that no $\lambda\mu\lambda$ 
travel is possible.

\bigskip

As the nullity of $s^2$ is, to the best of our knowledge, the only way to 
insure generically that the R-symmetric state has no tachyonic direction 
along $X$, we can then write a new conclusion:

\bigskip

\it VI. A generic renormalizable model has an R-symmetric 
supersymmetry-breaking vacuum for some non-negligible set of parameters 
if:

\begin{itemize}

\item it has a field $X$ with $R=2$ and no field with $R=-2$;

\item all other fields have R-charges which can be arranged in one or 
several of the following sequences or their subsequences:
\end{itemize}
\rm

$$
\begin{array}{ccc}
&&2\\
&\mu\swarrow&\\
\lambda\subset 0&&
\end{array};\quad\quad \mu\subset1\quad\hbox{OR}\quad
\begin{array}{ccccccc}
&&3&&&&5\\
&\mu\nearrow&&\lambda\diagdown&&\mu\swarrow\\
-1&&&&-3&&
\end{array} ;
$$

$$
\begin{array}{ccccccc}
&&R-2&&&&R\\
&\mu\nearrow&&\lambda\diagdown&&\mu\swarrow\\
4-R&&&&2-R&&
\end{array}.
$$

\bigskip

This is only a sufficient condition, but it seems indeed, as conjectured 
in \cite{Shih:2007av}, to make R symmetry preservation in 
supersymmetry-breaking systems something quite restrictive and R 
symmetry breaking something quite generic. Even when condition VI. is 
met, a suitable, but not fine-tuned, choice of parameters can very well 
make a tachyonic direction appear in the $\phi$s and spontaneously break 
R symmetry.

\section{Conclusion}

The main results of this study can be summarized as follows (table 
\ref{conclusions} gives a clearer picture of them):

\begin{itemize}

\item In a generic model with $n_0 \geq n_2$: if a vacuum breaks 
supersymmetry, then it breaks R symmetry (conclusion V.). There exists a 
symmetric vacuum (conclusion II.); there can exist R symmetry-breaking 
vacua, for instance if condition (\ref{NSconditions}) is met (in which 
case the vacua are supersymmetric), or not.

\item In a generic model with $n_0 < n_2$, in most cases there exists no 
supersymmetric vacuum (conclusions II. and III.), though some rare 
exceptions can be found (\ref{exception}). Moreover:

\begin{itemize}

\item If condition V. is not met in its entirety, there exists 
generically no R-symmetric vacuum either. There can exist a 
symmetry-breaking vacuum, or not (runaway case).

\item If condition V. is met, no definite conclusion can be reached as 
regards R symmetry breaking. Condition VI. guarantees the stability of the 
pseudomodulus potential, but not necessarily the absence of other 
tachyonic directions.

\end{itemize}

\end{itemize}

\bigskip

Phenomenologically, we would like to find simple conditions for both 
supersymmetry- and R symmetry-breaking. This is not really easy, since 
we have not found any conditions for the existence of vacua that 
preserve neither symmetry. Still, we can easily find nearly-sufficient 
conditions by looking for models that have neither R-symmetric nor 
supersymmetric vacua.

Those models must first have $n_2<n_0$. This guarantees that no 
R-symmetric state is supersymmetric. In most of these cases, therefore, 
supersymmetry shall be globally broken. Then we can forbid R-symmetric 
vacua by chosing models which do not meet condition V: either by using 
a superfield with R-charge $-2$, or by using R-charges that do not 
follow the pattern of condition V.

\bigskip

Such conditions (say models with $n_2>n_0$, $n_{-2}>0$) are not 
altogether sufficient: we must still check for a possible Nelson-Seiberg 
exception such as (\ref{exception}). We could not find any simple 
condition to avoid this type of case, but it looks quite exceptional. 
Overall, we found a reasonably large class of generic R-symmetric 
supersymmetric models where both supersymmetry and R symmetry are 
spontaneously broken. This agrees with the conclusions of 
\cite{Shih:2007av} and indicates that no fine tuning or \emph{ad hoc} 
features should be required in a physical model that accounts for the 
breaking of these unobserved symmetries.

\bigskip

Some directions should be further explored in times to come:

\begin{itemize}

\item In the case of the original \'O Raifeartaigh model, in the range 
of parameters where R symmetry seems to be broken by a tachyonic 
direction, it is in fact restored in a new vacuum. This is not 
altogether a surprise, since the tachyonic direction has R-charge zero 
and therefore locally preserves R symmetry, but it could show that R 
symmetry preservation is more robust in reality than it appears in this 
paper.

\item We have neglected the study of ordinary symmetries in our 
conclusions. A complete work should take them more explicitly into account 
in the definition of what is ``generic'' and what is not.

\item All gauges and attached D-terms have been let out of this work for 
simplicity's sake. Their inclusion could result in different 
conclusions. See \cite{Intriligator:2007py} for a recent study of 
supersymmetry-breaking in gauged models.

\end{itemize}

\section*{Annex I. Nelson-Seiberg exceptions}

This part illustrates the abstruse conditions (\ref{NSconditions}) under 
which R-symmetric Wess-Zumino models can circumvent the conclusion 
(IIIa.) of \cite{Nelson:1993nf} that forbids supersymmetric vacua when R 
symmetry is broken. Let us re-write these conditions: for some existing 
R-charge $R$,

$$
\left\{
\begin{array}{rcl}
n_2 \leq n_0 && \\
\forall E\subset \mathbb{R}^+, E\notin\{\emptyset,\{0\}\}, \sum_{\alpha
\in E} n_{2-\alpha R} &\leq& n_0 + \sum_{r\in F_E} \bar n_{r},
\end{array}\right.
$$

$F_E \equiv \{r; r/R>0, r/R$ summable up to $\alpha, \alpha-1,...$
for some $\alpha\in E\}$, $\bar n_r \equiv n_r - \delta_{r,R}$.

By ``summable up to...'', we mean that an element of $F_E$ must come in 
a possible sum of ratios of existing R-charges $r/R>0$ adding up to 
$\alpha, \alpha-1,...$.

We shall translate these conditions into simple inequalities in 
particular cases.

\bigskip

\emph{$\{-1,0,2\}$ models}

\bigskip

Consider a model with only fields of R charges $-1$, $0$ or $2$. Two 
possibilities arise for the breaking of R symmetry: either by $R=-1$ 
fields or by $R=2$ fields. Writing conditions (\ref{NSconditions}) for 
$R=-1$ only yields $n_2\leq n_0$, whereas $R=2$ yields in addition a 
condition $n_2+n_0+n_{-1} \leq n_0+n_2-1$ which is impossible. 
Therefore, for generic $\{-1,0,2\}$ models with $n_2\leq n_0, n_{-1}\geq 
1$, there are R symmetry-breaking supersymmetric vacua. Choosing for 
instance $n_{-1}=n_0=n_2=1$, the generic superpotential is:

\begin{equation}
W=\phi_2 f(\phi_0,\phi_2\phi_{-1}^2)
\end{equation}

so that supersymmetry conditions write:

\begin{eqnarray*}
f+\phi_2\phi_{-1}^2 \partial_2 f&=&0\\
\phi_2 \partial_1 f &=& 0\\
2\phi_2^2\phi_{-1} \partial_2 f &=&0.
\end{eqnarray*}

The solutions are generically given by $\phi_2=0, f(\phi_0,0)=0$, which 
lets $\phi_{-1}$ free to break R symmetry: this is indeed an exception 
to the Nelson-Seiberg result.

\bigskip

\emph{$\{-1,1,3\}$ models}

\bigskip

In these models, conditions (\ref{NSconditions}) amount for $R=-1$ to 
$n_{-1} \geq n_3 +1 $ and for $R=3$ to $n_3\geq n_{-1}+1$ (they are 
impossible for $R=1$, yielding $n_1\leq n_1-1$). Indeed, if we choose a 
$n_3=n_1=n_{-1}=1$ model, satisfying neither of these conditions, the 
generic superpotential writes:

\begin{equation}
W = \phi_3\phi_{-1} f(\phi_1\phi_{-1},\phi_3\phi_{-1}^3) + \phi_1^2 
g(\phi_1\phi_{-1}),
\end{equation}

and the only supersymmetric vacuum is at $\phi_3=\phi_1=\phi_{-1}=0$, 
which is the R-symmetric state (as we already knew from the fact that no 
R-charge 2 is present). The Nelson-Seiberg conclusion is here respected.

But if we consider the model $n_3=2,n_{-1}=1,n_1=0$, a generic 
superpotential is of the form:

\begin{equation}
W = X\phi f(X\phi^3,Y\phi^3) + Y\phi g(Y\phi^3)
\end{equation}

which has supersymmetric vacua for $\phi=0$, $Xf(0,0)+Yg(0)=0$, allowing 
for R symmetry breaking by a combination of the $R=3$ fields $(X,Y)$. A 
similar result is found for $n_3=1$, $n_{-1}=2$, where R symmetry is 
broken by fields with $R=-1$.

\section*{Annex II. Pseudomodulus one-loop potential}

Temporarily forgetting R symmetry, we consider a generic Wess-Zumino 
model with $N=n+1$ chiral superfields $X$ and $\phi^i$, $i=1\dots n$, 
with canonical K\"ahler potential and a renormalizable superpotential 
which, if it admits a supersymmetry-breaking (meta)stable vacuum, can 
always be written:

\begin{equation}
W=\xi X + \frac{1}{2}(\mu_{ij}+X \lambda_{ij})\phi^i \phi^j + 
\frac{1}{6}\gamma_{ijk}\phi^i\phi^j\phi^k
\end{equation}

The tree-level potential for the scalar fields $V_0 = 
|\partial_i W|^2$ has a flat direction along $X$, the scalar partner to
the goldstino, which therefore has zero mass at this level. However, the 
difference between the masses of the other particles along this 
(complex) line of vacua generates a degeneracy-lifting one-loop 
effective potential given by the Coleman-Weinberg formula:

\begin{equation}
V_1 = \frac{1}{64\pi^2}\Str\left(M^4 \log \frac{M^2}{\Lambda^2}\right)
\end{equation}

where $\Lambda$ is any fixed mass scale and $M$ is the tree-level mass 
matrix, given for the chosen form of the superpotential by:

\begin{equation}
M_0^2=\left(\begin{array}{cc}
\tilde\mu^\dagger\tilde\mu & \xi\lambda^\dagger \\
\xi\lambda & \tilde\mu \tilde\mu^\dagger\end{array}\right) , \;
M_{1/2}^2=\left(\begin{array}{cc}
\tilde\mu^\dagger\tilde\mu & 0 \\
0 & \tilde\mu \tilde\mu^\dagger\end{array}\right),
\end{equation}

where $\tilde\mu = \mu + X\lambda$. The trilinear interaction terms 
$\gamma_{ijk}$ do not come in this expression. $M_0^2$ must be positive 
in order for the vacuum to be stable with respect to oscillations in the 
$\phi$ directions.

In order to simplify notations, we shall use the following dimensionless 
parametrisation:

\begin{eqnarray*}
V_1&=&\frac{\Lambda^4}{64\pi^2} v \\
\mu &=& \Lambda a\\
\lambda &=& \frac{\Lambda^2}{\xi} b\\
X &=& \frac{\xi}{\Lambda} x \\
M^2 &=& \Lambda^2 P
\end{eqnarray*}

We shall moreover use a quantity $\epsilon$ which is $1$ for the scalar 
part and $0$ for the fermionic part. Thus $\epsilon^2=\epsilon$ and the 
supertrace of a matrix can be calculated as the term in $\epsilon$ in the 
trace of the matrix. We therefore have:

\begin{equation}
v=\Str(P^2\log P), \ 
P=\left(\begin{array}{cc}
\tilde a^\dagger \tilde a & \epsilon b^\dagger \\
\epsilon b & \tilde a \tilde a^\dagger\end{array}\right),
\end{equation}

where $\tilde a = a + xb$. The logarithm can be calculated, either by 
diagonalizing the matrix, a tiresome method that can hardly be exploited 
for general results, or by using holomorphic functional calculus:

\begin{equation}
v=\frac{1}{2i\pi} \oint dz\, z^2 \log z\, \Str\left[(z-P)^{-1}\right],
\end{equation}

where the contour encircles the whole spectrum of $P$ and (for 
logarithmic reasons) does not encircle zero\footnote{This is obviously 
possible for a discrete spectrum if and only if zero is not a part of 
the spectrum of P. But if it is, it can anyway be omitted since it does 
not contribute to $P^2\log P$. Note also that the spectrum of $P$ has to 
be positive in order for the tree-level potential to have a local 
minimum at that point.}.

We can easily calculate the supertrace of the inverse matrix, using the 
fact that $\tilde a$ and $b$ are symmetric matrices:

\begin{equation}
\Str\left[(z-P)^{-1}\right]=2\, \Str\left[\left(z-\tilde a^\dagger 
\tilde a - \epsilon b^\dagger (z-\tilde a \tilde a^\dagger)^{-1} 
b\right)^{-1} \right]
\end{equation}

This formula now enables us to make several interesting computations.

\bigskip\emph{Asymptotic behaviour}\bigskip

We can for instance look for the asymptotic behaviour as $x$ goes to 
infinity. Note that in every model (for any choice of $\lambda$, $\mu$), 
the bosonic mass matrix is positive as $|X|\rightarrow \infty$. That is 
obvious if we write for a given vector $\psi\equiv (\psi_1,\psi_2)$:

$$
\psi^\dagger M_0^2 \psi = \|(\mu+X\lambda)\psi_1\|^2 + 
\|(\mu+X\lambda)^\dagger \psi_2\|^2 + \xi \psi_1^\dagger \lambda^\dagger 
\psi_2 + \xi \psi_2^\dagger \lambda \psi_1
$$

This tends to positive infinity as $X\rightarrow\infty$ except if 
$\lambda\psi_1=\lambda^\dagger\psi_2=0$, in which case it is a constant, 
equal to $\|\mu\psi_1\|^2+\|\mu^\dagger \psi_2\|^2 \geq0$. Thus the 
one-loop potential is always well-defined for $|X|\rightarrow\infty$.

The norm of $P$ clearly goes to infinity as $|x|\rightarrow\infty$, so 
that we have to be careful of the contour we choose. We shall change 
variable in the integral by defining $w\equiv z/|x|^2$:

$$
v=\frac{|x|^4}{i\pi} \oint dw\, w^2 \log (w|x|^2) \,
\Str\left[w-\alpha^\dagger \alpha - \epsilon |x|^{-4} b^\dagger (w-\alpha 
\alpha^\dagger)^{-1} b\right]^{-1},
$$

where $\alpha \equiv b+x^{-1}a$. The poles of the supertrace no 
longer go to infinity as $x$ is great, so that we can choose a 
sufficiently large contour for the integration over $w$ and take the limit 
$|x|\rightarrow\infty$. As only the terms in $\epsilon$ are kept by the 
supertrace, it will have as its highest order term:

$$
v \sim \frac{\log |x|^2}{i\pi}\oint dw\, w^2\, \tr\left[(w-b^\dagger 
b)^{-2}b^\dagger (w-b b^\dagger)^{-1} b\right]\ .
$$

Using $(w-bb^\dagger)^{-1} b = b(w-b^\dagger b)^{-1}$, we calculate:

$$
\oint dw\, w^2\, \tr\left[(w-b^\dagger b)^{-2}b^\dagger (w-b 
b^\dagger)^{-1} b\right] = 2i\pi\,\tr(b^\dagger b)
$$

so that the asymptotic formula for $v$ is:

$$
v \sim 2\log |x|^2\, \tr(b^\dagger b)
$$

or, using the original physical quantities:

\begin{equation}
V_1(X) \sim \frac{|\xi|^2 \tr(\lambda^\dagger \lambda)}{32\pi^2}\log 
\left(\frac{|X|^2}{\Lambda^2} \right)
\end{equation}

This result agrees with what we found in the simple three-field case in 
\cite{Ray:2006wk}.

The coefficient is strictly positive for all supersymmetry-breaking 
models ($\xi\neq0$, $\lambda\neq0$), so that the potential always goes 
to infinity for infinite $X$. The modulus must then be either stabilized 
on a (meta)stable supersymmetry-breaking vacuum somewhere along the 
complex line $X$, or driven by negative masses of the $\phi$ fields to 
another, possibly supersymmetric, vacuum.

\bigskip\emph{Pseudomodulus one-loop mass calculation in 
R-symmetric cases}\bigskip

Coming back to our original discussion and following the work already 
done in \cite{Shih:2007av}, we shall consider an R-symmetric model. Then 
the potential for $X$ at $\phi_i=0$ should only depend on $|X|$ ---which 
already implies that $X=0$ is a local extremum of the potential---, so 
that we can choose $X$ real, which enables us to write:

$$
P=(A+xB)^2 + \epsilon B,\
A\equiv\left(\begin{array}{cc}
0 & a^\dagger\\
a& 0\end{array}\right),\
B\equiv\left(\begin{array}{cc}
0 & b^\dagger\\
b& 0\end{array}\right).
$$

The squared mass at the origin can then be found by writing:

\begin{eqnarray*}
\left.\frac{\partial^2 v}{\partial x^2}\right |_{x=0} &=& \frac{1}{2i\pi} 
\oint dz\, z^2\, \log z\, \Str\left[\frac{\partial^2}{\partial 
x^2}(z-P)^{-1}\right]\\
&=& \frac{1}{i\pi} \oint dz\, z^2\, \log z\\
&&\times \Str\left\{(z-A^2-\epsilon 
B)^{-2} \left[B^2 + \{A,B\}(z-A^2-\epsilon B)^{-1}\{A,B\}\right]\right\}
\end{eqnarray*}

Integrations by part give us:

$$
\oint dz\, z^2\, \log z B^2(z-P)^2 = \oint dz(2z\,\log z + z)\, B^2 
(z-P)^{-1}\\
$$

\begin{eqnarray*}
\oint dz\, z^2\, \log z\, \Str 
\left[(z-P)^{-2}\{A,B\}(z-P)^{-1}\{A,B\}\right]\\
= \oint \left(z\,\log z + \frac{z}{2}\right)\, 
\Str\left\{\left[\{A,B\}(z-P)^{-1}\right]^2\right\}
\end{eqnarray*}

so that

\begin{eqnarray*}
\left.\frac{\partial^2 v}{\partial x^2}\right |_{x=0} 
&=& \frac{1}{i\pi} \oint dz\, \left(z\, \log z+\frac{z}{2}\right) f(z),\\
f(z)&\equiv&\Str\left\{(z-A^2-\epsilon 
B)^{-1} \left[2B^2 + \{A,B\}(z-A^2-\epsilon B)^{-1}\{A,B\}\right]\right\}
\end{eqnarray*}

If we send $|z|$ to infinity, then $f(z)\sim 2z^{-3}\tr 
\left(B^4\right)$, where we have used the obvious nullity of the trace 
of any product of an odd number of matrices $A$, $B$. This shows that if 
the contour of the integral is chosen to be, firstly a very large circle 
from $R e^{-i\pi}$ to $R e^{i\pi}$, secondly the upper negative-real 
axis from $R e^{i\pi}$ to zero, thirdly the lower negative-real axis 
from zero to $R e^{-i\pi}$, then the circular part, behaving as 
$R^{-1}\log R$, tends to zero as $R\rightarrow \infty$ and the integral 
is equal to:

\begin{eqnarray*}
\left.\frac{\partial^2 v}{\partial x^2}\right |_{x=0} &=&
\frac{1}{i\pi} \int_0^{+\infty} dy\, \left(-y\, \log (y e^{i\pi}) - 
\frac{y}{2}\right)\, f(-y)\\
&&- \frac{1}{i\pi} \int_0^{+\infty} dy\, \left(-y\, \log (y e^{-i\pi}) - 
\frac{y}{2}\right)\, f(-y)\\
&=& -2\int_0^{+\infty} dy\, y f(-y)\\
&=& 4\int_0^{+\infty} dy\, y\, \Str\left\{(y+A^2+\epsilon 
B)^{-1} \right.\\
&& \times \left.\left[B^2 - \frac{1}{2}\{A,B\}(y+A^2+\epsilon 
B)^{-1}\{A,B\}\right]\right\}
\end{eqnarray*}

The mass of the pseudo-scalar $X$ is then $m_X^2 = 
\frac{\partial^2V_1}{2\partial X^2} = \frac{\Lambda^6}{128\pi^2\xi^2} 
\frac{\partial^2 v}{\partial x^2}$. We can define a dimensionless mass 
$q$ by $m_X^2 \equiv \frac{\Lambda^6}{32\pi^2\xi^2} q^2$, so that:

\begin{equation}
\begin{array}{rcl}
\displaystyle
q^2 &=& \int_0^{+\infty} dy\, y\, \Str\left\{(y+A^2+\epsilon 
B)^{-1} \right.\\
&& \times \left.\left[B^2 - \frac{1}{2}\{A,B\}(y+A^2+\epsilon 
B)^{-1}\{A,B\}\right]\right\}.
\end{array}
\end{equation}

This proves that equation (2.12) in \cite{Shih:2007av} is exact in the 
limit $\Lambda\rightarrow \infty$.

Following that paper, we then define a matrix $F(y)\equiv (y+A^2)^{-1} B$, 
allowing us to write:

\begin{eqnarray*}
q^2 &=& \int_0^{+\infty} dy\, y\, \Str\left\{(1+\epsilon F(y))^{-1} 
F(y)^2\left(y+A^2\right) \right.\\
&&- \left.\left[F(y)(1+\epsilon F(y))^{-1} A\right]^2
- F(y)^2 \left[(1+\epsilon F(y))^{-1} A\right]^2\right\}.
\end{eqnarray*}

From the block-antidiagonal structure of $F$ and $A$ we can deduce the 
following identities:

\begin{eqnarray*}
\tr\left[F^2(1+F)^{-1}\right] &=&\tr\left[F^2(1-F^2)^{-1}\right]\\
\tr\left[F^2(1+F)^{-1}A^2\right] &=&\tr\left[F^2(1-F^2)^{-1}A^2\right]\\
\tr\left\{\left[F(1+F)^{-1}A\right]^2\right\} &=& 
\tr\left\{\left[F^2(1-F^2)^{-1}A\right]^2 + \left[F(1-F^2)^{-1} 
A\right]^2 \right\}\\
\tr\left\{F^2 \left[(1+F)^{-1}A\right]^2\right\} &=& \tr\left\{F^2 
\left[(1-F^2)^{-1}A\right]^2 + F^2 \left[F(1-F^2)^{-1}A\right]^2\right\} 
\end{eqnarray*}

We now use R symmetry to eliminate the second term in the last two 
identities: the matrix $F^2$ connects elements with the same R-charge, 
but  $F$ connects a charge $R$ with $-R$ and $A$ a charge $R$ with 
$2-R$, so that $F(1-F^2)^{-1}A$ connects a charge $R$ with $2+R$: the 
trace of the two expressions involving this matrix is therefore zero. 
The same can be said of $\tr(FAFA)$, so that:

\begin{equation}
\begin{array}{rcl}
\displaystyle
q^2 &=& \int_0^{+\infty} dy\, \tr\left\{y^2F(y)^4(1-F(y)^2)^{-1}\right. \\
&& \left.- 2y\left[F(y)^2(1-F(y)^2)^{-1}A\right]^2 \right\}
\end{array}
\end{equation}

This formula was first found in \cite{Shih:2007av} (2.14). Some of its 
properties will be easier to prove if we replace $F$ by the following 
matrix:

\begin{equation}
G(y)\equiv (y+A^2)^{-1/2} B (y+A^2)^{-1/2},
\end{equation}

where the square root is (uniquely) defined in the sense of positive 
definite matrices, $(y+A^2)$ being, for $y>0$, such a matrix. Note that 
$G$ is hermitian. The cyclicity of the trace enables us to use 
indifferently $F$ or $G$ in a trace formed of $F$ and $A$ only, so that:

\begin{equation}
\left\{
\begin{array}{rcl}
q^2 &=& r^2-2s^2,\\
r^2 &\equiv& \int_0^{+\infty} dy\, y^2\, \tr\left[G(y)^4(1-G(y)^2)^{-1} 
\right],\\
s^2 &\equiv& \int_0^{+\infty} dy\, y \,
\tr\left\{\left[G(y)^2(1-G(y)^2)^{-1}A\right]^2 \right\}.
\end{array}
\right.
\end{equation}

We shall now verify that both $r^2$ and $s^2$ are positive and can only 
be zero under certain conditions.

First notice that the bosonic mass matrix $A^2+B$ is positive, meaning, 
given the expression of these matrices, that:

$$
\forall \psi_1,\psi_2,\, \|a\psi_1\|^2 + \|a^\dagger 
\psi_2\|^2+2\Re(\psi_2^\dagger b \psi_1) \geq0.
$$

The transformation $b\rightarrow -b$ is then just equivalent to a change 
of variables $\psi_1\rightarrow -\psi_1$, so that $A^2-B$ is positive too. 
We can write more generally:

$$
\forall y>0,\, y+A^2\pm B >0
$$

As $y+A^2$ is positive definite and as:

$$1\pm G(y) = (y+A^2)^{-1/2}(y + A^2 \pm B)(y+A^2)^{-1/2},$$

we have $1\pm G(y)>0,$ so that:

$$\forall y>0,\, 1-G(y)^2 >0.$$

$G(y)$ being hermitian, $G(y)^2$ is obviously positive and so is 
therefore $G(y)^2(1-G(y)^2)^{-1}G(y)^2$. From this we deduce that $r^2$ 
is always positive and is zero only if $G(y)=0$, that is, 
$B=0=\lambda$, which is the utterly uninteresting case where the 
supersymmetry-breaking field $X$ is decoupled from all the rest.

\bigskip

The trace that appears in the expression of $s^2$ can be written:

\begin{equation}
\tr\left\{\left[(1-G(y)^2)^{-1/2}G(y)A\, 
G(y)(1-G(y)^2)^{-1/2}\right]^2\right\} \geq0,
\end{equation}

so that $s^2\geq0$. $s^2$ is zero if and only if:

\begin{eqnarray*}
\,&\,& \forall y>0,\, G(y)A\, G(y)=0\\
&\Leftrightarrow & \forall y>0,\, BA(A^2+y)^{-1}B=0\\
&\Leftrightarrow & \forall k\geq0,\, BA^{1+2k}B=0\\
&\Leftrightarrow & \forall k\geq0,\, b(a^\dagger a)^k a^\dagger b=0\\
&\Leftrightarrow & \forall k\geq0,\, \lambda(\mu^\dagger \mu)^k 
\mu^\dagger \lambda=0.
\end{eqnarray*}

\textbf{Acknowledgements}

\bigskip

My thanks first go to Costas Bachas, whose advice was of considerable 
worth in the course of this work. I repeat here my debt to the 
stimulating article of David Shih and thank Nathan Seiberg for his kind 
encouragements on account of my previous work.


\begin{thebibliography}{00}

\bibitem{Wess:1973kz}
  Julius~Wess and Bruno~Zumino,
  \emph{A Lagrangian Model Invariant Under Supergauge Transformations},
  Phys.\ Lett.\  B {\bf 49} (1974) 52.

\bibitem{Raifeartach:1975pr}
  Lochlainn~\'O Raifeartaigh,
  \emph{Spontaneous Symmetry Breaking For Chiral Scalar Superfields},
  Nucl.\ Phys.\  B {\bf 96} (1975) 331.

\bibitem{Witten:1981kv}
  Edward~Witten,
  \emph{Mass Hierarchies In Supersymmetric Theories},
  Phys.\ Lett.\  B {\bf 105} (1981) 267.

\bibitem{Witten:1982df}
  Edward~Witten,
  \emph{Constraints On Supersymmetry Breaking},
  Nucl.\ Phys.\  B {\bf 202} (1982) 253.

\bibitem{Affleck:1983vc}
  Ian Affleck, Michael Dine and Nathan Seiberg,
  \emph{Dynamical Supersymmetry Breaking In Chiral Theories},
  Phys.\ Lett.\  B {\bf 137} (1984) 187.

\bibitem{Nelson:1993nf}
  Ann~E.~Nelson and Nathan~Seiberg,
  \emph{R symmetry breaking versus supersymmetry breaking},
  Nucl.\ Phys.\  B {\bf 416} (1994) 46.

\bibitem{Barbier:2004ez}
  R.~Barbier {\it et al.},
  \emph{R-parity violating supersymmetry},
  Phys.\ Rept.\  {\bf 420} (2005) 1
  [arXiv:hep-ph/0406039].

\bibitem{Intriligator:2006dd}
  Kenneth~A.~Intriligator, Nathan~Seiberg and David~Shih,
  \emph{Dynamical SUSY breaking in meta-stable vacua},
  JHEP {\bf 0604} (2006) 021
  [arXiv:hep-th/0602239].

\bibitem{Ray:2006wk}
  S\'ebastien~Ray,
  \emph{Some properties of meta-stable supersymmetry-breaking vacua in 
Wess-Zumino models},
  Phys.\ Lett.\  B {\bf 642} (2006) 137.

\bibitem{Intriligator:2007cp}
  Kenneth~A.~Intriligator and Nathan~Seiberg,
  \emph{Lectures on Supersymmetry Breaking},
  arXiv:hep-ph/0702069.

\bibitem{Shih:2007av}
  David~Shih,
  \emph{Spontaneous R-symmetry breaking in O'Raifeartaigh models},
  arXiv:hep-th/0703196.

\bibitem{Intriligator:2007py}
  Kenneth~A.~Intriligator, Nathan~Seiberg and David~Shih,
  \emph{Supersymmetry Breaking, R-Symmetry Breaking and Metastable Vacua},
  arXiv:hep-th/0703281.

\bibitem{Ferretti:2007ec}
  Luca~Ferretti,
  \emph{R-symmetry breaking, runaway directions and global symmetries in
  O'Raifeartaigh models},
  arXiv:0705.1959 [hep-th].

\end{thebibliography}
\end{document}